\documentstyle[12pt,aasms4]{article}

\lefthead{Fargion, Mele and Salis}
\righthead{Ultrahigh energy neutrino scattering} 
\begin{document}


\title{Ultrahigh energy neutrino scattering onto relic light\\
neutrinos in galactic halo as a possible source\\
of highest energy extragalactic cosmic rays}

\author{D. Fargion\altaffilmark{1,2}, B. Mele\altaffilmark{1} and A. Salis}
\affil{Dipartimento di Fisica, Universita' degli Studi di Roma\\ 
''La Sapienza'', P.le A. Moro 2, 00185 Roma, Italy.}

\authoremail{Daniele.Fargion@roma1.infn.it}

\altaffiltext{1}{INFN, Roma1, Italy}
\altaffiltext{2}{Technion Institute, Eng. Faculty, Haifa, Israel}

\begin{abstract}

The diffuse relic neutrinos with light mass are transparent to Ultrahigh 
energy (UHE) neutrinos at thousands EeV, born by photoproduction of 
pions by UHE protons on relic 2.73 K BBR radiation and originated in 
AGNs at cosmic distances. However these UHE $\nu$s may interact with those 
(mainly heaviest 
$\nu_{\mu_r}$, $\nu_{\tau_r}$ and respective antineutrinos) clustered into HDM 
galactic halos. UHE photons or protons, secondaries of $\nu\nu_r$ scattering, 
might be 
the final observed signature of such high-energy chain reactions and may be 
responsible of the highest extragalactic cosmic-ray (CR) events. The 
chain-reactions conversion efficiency, ramifications and energetics are 
considered 
for the October 1991 CR event at 320 EeV observed by the Fly's Eye 
detector in Utah. These quantities seem compatible with the distance, direction 
and power (observed at MeV gamma energies) of the Seyfert galaxy MCG 8-11-11. 
The $\nu\nu_r$ interaction probability is favoured by at least three order 
of magnitude with respect to a direct $\nu$ scattering onto the Earth 
atmosphere. 
Therefore, it may better explain the extragalactic origin of the puzzling 320 
EeV event, while offering indirect evidence of a hot dark galactic halo of 
light ($\it{i.e.}$, $m_\nu\sim$ tens eV) neutrinos, probably of tau flavour.  
\end{abstract}


\section{Introduction}

The highest energy CR, with $E>10^{19}~eV$, (excluding neutrinos) 
are severely bounded to 
nearby cosmic distances by the opacity of the $2.73~K$ BBR (GZK cutoff) 
\markcite{gre66,zk66,gos66} (Greisen 1966; Zat'sepin, Kuz'min 1966; Gould, 
Schreder 1966) 
as well as by the extragalactic radiobackground opacity \markcite{cba70, 
prb96} (Clark, Brown, Alexander 1970; Protheroe, Biermann 1996). The 
Inverse Compton Scattering ($e^{\pm}\gamma_{BBR}\rightarrow e^{\pm}\gamma$, 
$p\gamma_{BBR}\rightarrow p\gamma$, \markcite{lon94,fks97} (Longair 1994; 
Fargion, Konoplich, Salis 1997)), the photopair production at 
higher energies ($p\gamma_{BBR}\rightarrow pe^+e^-$, $\gamma\gamma_{BBR}
\rightarrow e^+ e^-$) and, most importantly, the photoproduction of 
pions ($p\gamma_{BBR}\rightarrow p+N\pi$, 
$n\gamma_{BBR}\rightarrow n+N\pi$,...) \markcite{lon94} (Longair 1994) 
constrain a hundred-EeV CR to few Mpc for the characteristic pathlenghts of 
either charged CR (protons, nuclei), as well as neutrons and photons. The most 
energetic CR event of 320 EeV, at the Utah Fly's Eye detector, 
keeps the primeval source direction within few degrees, even if it was a 
charged one, because 
of its high magnetic rigidity \markcite{bir94} (Bird et al. 1994). 
However, no nearby ($<60~Mpc$) source candidate 
(AGN, QSO) has been found up to now in the arrival direction error box. 
Therefore, presently, there is no reasonable explanation for the 320 EeV 
event, in case its origin is extragalactic. An alternative solution to this 
puzzle, 
based on an extraordinary extragalactic magnetic field (whose coherence lenght, 
ranging the largest astrophysical distances, might be able to bend the CR 
trajectory from off-axis nearby potential sources, like M82 and Virgo A 
\markcite{els95} (Elbert, Sommers 1995)) is improbable \markcite{met97} 
(Medina-Tanco 1997). A $\it{local}$ (galactic halo) origin of the cosmic ray,
is unpopular because of the lack of known processes able to 
accelerate cosmic rays in small galactic objects as SN remnants or jets up to
such high energies. Direct
cosmic neutrinos reaching the Earth atmosphere, while 
able to reach the Earth from any cosmological distance, are unable to produce 
the observed shower of the 320 EeV event \markcite{els95} (Elbert, Sommers 
1995). Exotic sources of the EeV CR as 
monopole decay or topological defect annihilation have also been considered 
\markcite{els95} (Elbert, Sommers 1995). However such models do not provide 
any detailed prediction (no one knows the primordial monopoles density) and 
they seem just $\it{a~posteriori,~ad~hoc}$ solutions. Here we want to address a 
more 
conventional solution based on the widely accepted assumption that neutrinos 
have a light mass and therefore they could cluster into large galactic halos, 
where they play an important role as $\it{hot~dark~matter}$. Their number 
density, cross sections and 
halo sizes are large enough to produce, by $\nu\nu$ electroweak $W^{\pm}$/Z 
bosons exchange, secondaries inside the galactic halo, mainly photons 
by $\pi^o$ decay and protons, which could be the source of the observed 320 EeV 
event \markcite{fas97} (Fargion, Salis 1997).

\section{Relic neutrino clustering}

Let us consider the UHE $\nu$ scattering onto light relic $\nu$'s in the 
galactic halo. In principle, any neutrino flavor may be involved either as a 
source or as a target for the present 
process. Let us label by $\nu$ the hitting high-energy neutrino and by $\nu_r$ 
the target relic one. UHE electronic neutrinos may derive from neutron decay 
in flight, 
or at lower energies from muon decay. Muonic neutrinos may be born as 
secondary in pion decay. Tau neutrinos may occur if the primary cosmic rays are 
generated in hadronic interactions \markcite{dal96} (Dar, Laor 1996) 
or if some neutrino flavor mixing occurs \markcite{far97} (Fargion 1997). 
The target neutrinos, relic of early Universe, are clustered around the 
galactic halo. All the neutrino flavours are born at nearly the same cosmic 
homogeneous BBR density $n_{\nu}\sim (4/11) n_{\gamma}$. Neutrinos with a light 
mass (few or tens eV) must condense around the galactic halos because of their 
earliest decoupling from 
thermal equilibrium, and because of the mutual barion-neutrino multifluid 
gravitational clustering during the galaxy formation \markcite{zel80,far83} 
(Zel'dovich 1980; Fargion 1983). Because of the 
neutrino mass role in defining the early Jeans instability, the free streaming 
mass and the halo size, the heavier (tau or mu) neutrino halos are expected 
to be more clustered and dense than a lighter (electronic) halo.\newline 
Let us review 
in more detail the role and the origin of such HDM neutrinos in galactic halos 
and their role in the interaction with the UHE neutrinos. In the early 
Universe, the thermal equilibrium provides the most efficient source 
of the present neutrino density, whose relic number may be easily derived by 
entropy conservation. The MeV neutrino relics decouple from the thermal bath 
of photons during the first second of the Universe life, as soon as electron 
pairs 
(which play the role of catalizators for the neutrino equilibrium) annihilate, 
and heat the relic photons temperature by an extra factor $(11/4)^{1/3}$ with 
respect to the 
neutrino one. The cosmological neutrino density, for each flavor and state, 
stems directly from the previous factor and becomes $n_{\nu}=(4/11) 
n_{\gamma}\simeq 108~cm^{-3}$. Charged currents keep the electronic neutrinos 
in thermal 
equilibrium, while the muonic and tauonic ones are kept in equilibrium by the 
slightly less efficient neutral currents. The cosmic expansion $\it{cools}$ the 
ultrarelativistic neutrino and, as soon as it reaches the 
non-relativistic regime $\kappa_B T_{\nu}\simeq m_{\nu} c^2$, allows the 
collisionless neutrino fluid to grow its density contrast \markcite{far83} 
(Fargion 1983). The barion density perturbations are meanwhile 
smeared out by photons up to the Silk size and mass, at a later 
recombination epoch ($z\sim 1500$). Once the barions decouple from radiation, 
their density contrast $\delta\rho_B/\rho_B$ may grow around the primordial 
neutrino gravitational seeds. Moreover, the barions may dissipate (by 
radiation) their gravitational energy, leading to faster non-linear 
gravitational galaxy formation. At 
this stage, the massive neutrinos are at their time sinked by the barionic 
galactic growing (gravitational) potential, and they finally fill up an 
extended hot dark galactic halo, $\it{i.e.}$, a HDM halo. We 
apply a simple adiabatic approximation to evaluate the present neutrino number 
density in those halos \markcite{zel80} (Fargion et al. 1995, 1996). 
Therefore, the final neutrino number density in the galactic halo $n_{\nu_r}$ 
is enhanced by a factor $\rho_{GB}/\rho_B\sim 10^5\div 
10^7$, where $\rho_{GB}$ and $\rho_B$ label respectively the present barionic 
mass density in inhomogeneous galaxies and in average cosmic 
media. In the present work we assume $n_{\nu_r}= 10^{7\div 9}~cm^{-3}$. The two 
order of magnitude of uncertainty window is related to our 
ignorance of the exact cosmic barion density and galactic (luminous and dark) 
mass density, as well as to the neutrino barion clustering efficiency. The 
consequent extended neutrino halo, related to the combined free-streaming 
length and characteristic Jeans wavelenght of the two fluids \markcite{far83} 
(Fargion 1983) is $l_g\simeq 300~Kpc\sim 10^{24}~cm$. The characteristic mass
density needed to solve the galactic dark matter problem is $\rho_{o c}\sim
0.3~GeV~cm^{-3}$ (Fargion et al. 1995). The corresponding allowed mass
density for the relic neutrino halo, for the two extreme clustered value we
assumed above, is $\rho(r)_{o \nu}\sim 0.1/(1+(r/a)^2) (m_\nu/10~eV)\cdot 10^
{0\div 2}~GeV~cm^{-3}$ where $a\sim 10~Kpc$ well within the critical value
$\rho_{o c}$.

\section{Neutrino-neutrino interaction}

Now, let us examine the processes that can occur in the 
interaction of the UHE neutrino with the relic one (see also \markcite{rou} 
Roulet 1993). The two 
main channels involve a $W^\pm$ or a $Z^o$ exchange via the reactions 
$\nu_\mu\nu_{\tau_r}\rightarrow \mu\tau$ and 
$\nu_\mu\nu_{\mu_r}\rightarrow hadrons$, respectively. 
For a $\nu\nu_r$ interaction mediated in the $t$-channel by the W exchange, 
the asymptotic cross section reaches a plateau of nearly constant value 
when $s\rightarrow\infty$. On the other hand, for a $\nu\nu_r$ interaction 
mediated in the $s$-channel by the Z exchange, a peculiar peak in the cross 
section occurs due to the resonant Z production at $s= M_Z^2$. However, this 
occurs for a very narrow and fine-tuned window of energy, and a neutrino 
mass $m_\nu\sim 4~eV (E_\nu /10^{21}~eV)^{-1}$. This resonance for massive 
light cosmological neutrinos is analogous to the well known one in $\nu_e e^-
\rightarrow W^-$ \markcite{gla60} (Glashow 1960; Berezinsky, Gazirov 1977). 
Here, we just notice this possibility, but we do not assume the lucky 
coincidence. The exact cross section for the $\nu_\mu\bar{\nu}_{\tau_r}$ 
(and charge conjugated $\bar{\nu}_\mu\nu_{\tau_r}$) interaction via a W 
exchange in the $t$-channel, 
neglecting the neutrino masses, is 
\begin{equation}
\sigma_W(s)= \sigma_{asym}\frac{A(s)}{s}\left\{ 
1+\frac{M_W^2}{s}\Bigg[ 2-\frac{s+B(s)}{A(s)}\ln\Bigg(\frac{B(s)+A(s)}{B(s)-
A(s)}\Bigg)\Bigg]\right\}
\end{equation}
where $\sqrt{s}$ is the center of mass energy, the functions A(s), B(s) are 
defined as
\begin{equation}
A(s)=\sqrt{[s-(m_\tau+m_\mu)^2][s-(m_\tau-m_\mu)^2]}~~~;~~~
B(s)=s+2M_W^2-m_\tau^2
\end{equation}
and 
\begin{equation}
\sigma_{asym}=\frac{\pi\alpha^2}{2\sin^4\theta_W M_W^2}\simeq~108.5~pb
\end{equation}
where $\alpha$ is the fine structure constant and $\theta_W$ the Weinberg 
angle. $\sigma_{asym}$ is the asymptotic behaviour of the cross section in the 
ultrarelativistic limit 
\begin{equation}
s\simeq 2 E_\nu m_\nu= 2\cdot 10^{23} (E_\nu /10^{22}~eV)(m_\nu / 10~eV)~eV^2
\gg M_W^2~~~~~.
\end{equation}
On the other hand, the interaction of neutrinos of the same flavour can occur 
via a Z exchange in the $s$-channel ($\nu_i\bar{\nu}_{i_r}$ and charge 
conjugated). The cross section for hadron production in 
$\nu_i\bar{\nu}_i\rightarrow Z^*\rightarrow hadrons$ is 
\begin{equation}
\sigma_Z(s)=\frac{8\pi s}{M_Z^2}\frac{\Gamma(Z^o\rightarrow invis.)
\Gamma(Z^o\rightarrow hadr.)}{(s-M_Z^2)^2+M_Z^2 \Gamma_Z^2}
\end{equation}
where $\Gamma(Z^o\rightarrow invis.)\simeq 0.5~GeV$, $\Gamma(Z^o\rightarrow 
hadr.)\simeq 1.74~GeV$ and $\Gamma_Z\simeq 2.49~GeV$ are respectively the 
experimental Z 
width into invisible products, the Z width into hadrons and the Z full 
width \markcite{pdg} (Particle Data Group, 1996). 
Apart from the narrow Z resonance peak at $\sqrt{s}= M_Z$, the 
asymptotic behaviour is proportional to $1/s$ for $s\gg M_Z^2$. For 
energies $\sqrt{s}> 2 M_W$, one has to include the additional channel of W pair 
production, $\nu_i\bar{\nu}_{i_r}\rightarrow W^+W^-$. The corresponding cross 
section is \markcite{enq89} (Enquist, Kainulainen, Maalampi 1989) 
\begin{equation}
\sigma_{WW}(s)=\sigma_{asym}\frac{\beta_W}{2s}\frac{1}{(s-M_Z^2)}\left\{4 L(s)
\cdot C(s)+D(s)\right\}
\end{equation}
where $\beta_W=(1-4 M_W^2/s)^{1/2}$ and the functions $L(s)$, $C(s)$, $D(s)$ 
are defined as
\begin{displaymath}
L(s)=\frac{M_W^2}{2\beta_W s}\ln\Big(\frac{s+\beta_W s-2 M_W^2}{s-
\beta_W s-2 M_W^2}\Big)
\end{displaymath}
\begin{equation}
C(s)=s^2+s(2 M_W^2-M_Z^2)+2 M_W^2(M_Z^2+M_W^2)
\end{equation}
\begin{displaymath}
D(s)=\frac{1}{12 M_W^2 (s-M_Z^2)}\times
\end{displaymath} 
\begin{displaymath}
\times\Big[ s^2(M_Z^4-60 M_W^4-4 M_Z^2 M_W^2)
+20 M_Z^2 M_W^2 s ( M_Z^2+2 M_W^2)-48 M_Z^2 M_W^4(M_Z^2 + M_W^2) \Big]~.
\end{displaymath}
The asymptotic behaviour of this cross section is proportional to
$(M_W^2/s)\ln{(s/M_W^2)}$ for $s\gg M_Z^2$. In fig.1 we show the three cross 
sections eq.1, eq.5, eq.6 as functions of $\sqrt{s}$.
\placefigure{fig1} 
Of course, in our approach we assume a relic neutrino mass of at 
least a few eV's. A too light neutrino would hardly cluster in galactic 
halo. Hence, a suppression factor 1/2 arises in the cross sections since in 
the non 
relativistic rest frame of the gravitationally bounded neutrinos the helicity 
of the particle is undefined. This means that the relic neutrino may be found 
in either left or right handed polarization states. Consequently, the 
interaction may be either left-handed, $\it{active}$, or right-handed, 
$\it{sterile}$, leading to the above suppression factor. Maiorana neutrinos,
which we do not consider here, are insensible to such suppression.\newline
Cosmic distances $l_c\simeq H^{-1} c\simeq 10^{28}~cm$ are transparent 
to UHE neutrinos even for massive diffused neutrinos (with interaction 
probability $P_c\simeq 
\sigma_{\nu\nu_r} n_\nu l_c\sim 10^{-4}$). However, denser extended neutrino 
halos are a more efficient calorimeter. The interaction probability via W 
exchange ($t$-channel) is $P_g\simeq \sigma_{\nu\nu_r} n_{\nu_r} 
l_g\sim 10^{-3}\div 10^{-1}$, $\it{i.e.}$ at least four 
order of magnitude larger than the corresponding interaction probability of 
UHE $\nu$'s ($E_\nu\sim 10^{21}~eV$) in terrestrial atmosphere 
($P_a=\sigma_{\nu\nu} n_{atm} l_{atm}\sim 10^{-5}$, \markcite{gan96} Gandhi 
et al. 1996) with an additional suppression 
factor ($\sim 10^{-2}$) due to the high altitude where the 320 EeV cosmic ray 
event took place.\newline
The main reaction chains, from the primary proton down to the 
final 320 EeV photons or protons, via neutrino-neutrino interactions, are 
described and 
summarized in tables 1, 2, 3, 4, 5 respectively for the $W^{\pm}$
($t$ channel), the $Z^o$ ($s$ channel) and the 
$\nu\bar{\nu}\rightarrow W^+W^-$ channel for pion production and the $Z^o$ 
channel and 
the $\nu\bar{\nu}\rightarrow W^+W^-$ channel for proton production. 
The final photons are mainly $\pi^o$ decay relics from either the $\tau^\pm$, 
the $Z^o$ or the $W^\pm$ secondaries born in $\nu\nu_r$ interactions in 
the galactic halo. The final protons\footnote{We remind the reader that a 
vector boson 
hadronic decay generates protons as well as antiprotons. For our purposes the 
two kind of particles are equally suitable and we refer to both of them as 
protons.} are among the secondaries of the $Z^o$ 
or $W^\pm$ hadronic decay.

\section{The chain reactions leading to the final photons}

There are at least two main sources of UHE neutrinos: photoproduction of pions 
by interaction of protons and neutrons on the BBR photons \markcite{lon94} 
(Longair 
1994) and $pp$ scattering \markcite{dal96} (Dar, Laor 1996). The neutrino 
UHE secondaries, that are able to survive from cosmic 
distances up to the neutrino galactic halo, are the first born muonic and 
antimuonic from the $\pi^\pm$ decay and the secondary born from the subsequent 
$\mu^\pm$ decay.\newline 
Let us first consider the three different chains of reactions giving rise to 
final photons. They refer respectively to the interaction of UHE 
$\nu$s with relic $\nu$s via $W^\pm$ exchange ($t$-channel, table 1), $Z^o$ 
exchange ($s$-channel, table 2) and $\nu\bar{\nu}\rightarrow W^+ W^-$ 
scattering (table 3).\newline
All the reactions in tables 1,2,3 assume 
an UHE $\nu$ born in photoproduction of pions from primary protons (CR protons) 
onto BBR photons (through either $p\gamma\rightarrow p+N \pi$ or 
$p\gamma\rightarrow n+ N \pi$, where the pion multiplicity is $N\geq 2$). 
The primordial proton energy is calibrated according to the 
final CR photon energy of $320~EeV$ and the different efficiencies of the 
chains themselves. In other words, this means that we started with the energy 
of 320 EeV
of the observed particle and we walked back the chain from the end to the
beginning according to the chains we have proposed in order to obtain the
fundamental input parameter: the primordial proton energy that obviously
depends on the chain.\newline
For the $W^\pm$ $t$-channel (table 1), this initial proton energy is 
relatively small, and the pions, produced by photoproduction, are therefore 
few ($\sim 2\pi$).\newline 
For the $Z^o$ $s$-channel (table 2), the energy is huge, so the corresponding 
photoproduced pion number is very large ($\sim 12\pi$).\newline
For the $W^+ W^-$ 
production (table 3) the energy is quite high and the number of photoproduced 
pions is not too small ($\sim 10\pi$)\footnote{The $\pi$ multiplicities have 
been estimated by assuming the 
scaling law $N\propto s^{1/4}$ and the fact that the charged pions are 2/3 
of the total number.}. The probability, multiplicity and secondary energies are 
easily derived as in tables 1,2 and 3.\newline
The successive pion decay $\pi^\pm\rightarrow \mu^\pm + 
\nu$ (eq. 2a,2b,2c in tables 1,2,3) and muon decay $\mu^\pm\rightarrow e^\pm 
\nu_e \nu_{\mu}$ are the main sources of (muonic) UHE neutrinos. 
These $\it{blind}$ neutral particles travel through cosmic distances without 
interacting, and then reach our (neutrino) galactic halo.\newline
Now, the main 
reaction differentiating the three possible chains in tables 1,2,3, is the 
UHE $\nu$ scattering onto relic $\nu$s in the hot dark matter halo 
(eq. 3a,3b,3c in tables 
1,2,3). The incoming hitting neutrinos are of muonic nature, while the target 
relic ones, because of the heavier mass, are preferentially tauonic. 
Indeed, the gravitational clustering of relic cosmic neutrinos in the HDM 
halo is 
more efficient for heavier (and slower) $\nu_{\tau_r}$, $\bar{\nu}_{\tau_r}$ 
\markcite{zel80,far83,fareta96} (Zel'dovich 1980; Fargion 1983; Fargion et 
al. 1996). For instance, in analogy to the neutron stars, in an ideal 
$\it{neutrino~star}$ the degenerated equation for the HDM galactic number 
density of neutrinos grows as $m_\nu^3$. As one may easily notice in 
figure 1, where the different cross sections are plotted, the $W^\pm$ 
$t$-channel interaction cross section (table 1, eq.1) reaches an asymptotic 
plateau, while the $Z^o$ $s$-channel cross section (table 2, eq.5) exhibits an 
interesting resonance at $s\sim M_Z^2$ (although, as discussed before, for a 
very narrow $m_\nu$ mass range), and decreases at higher energies. Also, the 
$W^+ W^-$ production, while of some importance at $E_{cm}\sim 2 M_W^2$ 
(where it is comparable with the $W^\pm$ $t$-channel), is also falling as 
ln(s)/s.\newline
The final secondaries of the $Z^*$, $W^\pm$ and $\tau^\pm$ decays 
are (according to the corresponding multiplicity and probability) source of 
$\pi^o$'s whose final photons may be the observed highest energetic cosmic 
rays.\newline 
We notice that the off-shell $Z^*$ decay channel (table 2, eq.5) produces a 
large population of 
$\pi^o$ secondaries ($\sim 20-21\pi^o$)\footnote{The $\pi$'s multiplicity has 
been obtained from the first equation in section 6.2 in \markcite{sch} 
(Schmelling 1995) and by assuming that the fraction of charged particle is 
nearly conserved at energies higher than $\sqrt{s}=M_Z$. Finally, we assumed 
that all the secondaries get about the same amount of energy from the 
off-shell $Z^*$ decay.}. 
The $W^+ W^-$ channel (table 3, eq.6) leads also to a large number of 
$\pi^o$ ($\sim$ 8 $\pi^o$)\footnote{The $\pi$'s multiplicity has been 
obtained by supposing that the $W^\pm$ hadronic decays are similar to the 
$Z^o$ ones. The $Z^o$ hadron multiplicities can be found in \markcite{kno} 
(Knowles et al. 1996).}. The 
most favourable chain, for energetics, is the $W^\pm$ exchange in the 
$t$-channel that 
occurs via the $\tau^\pm$ hadronic decay. The three chains with the 
corresponding 
primordial energy and total probabilities are summarized in tables 1,2,3 and 
will be further discussed in the conclusions. Even if we carried on the 
computations keeping split the two $\nu_\mu$ branches stemming from the $\pi$ 
and $\mu$ decays for sake of completeness, now we will refer to the specified 
chain probability as the sum of the two previous ones.\newline 
Let us remark that in the UHE neutrino-neutrino scattering the electron 
neutrino coming from the $\mu$ decay can have an important 
role\footnote{Another way to get electronic antineutrino is from the 
neutron beta decay in flight. These $\bar{\nu}_e$s are extremely energetic 
and very good candidate for energy transfer. Unfortunately, the free neutron 
decay, usually important at UHE energy of the order of $E_n\sim 10^{20}~eV$, 
becomes less and less competitive at 
higher Lorentz boost. Indeed, at the energies higher than $10^{21}~eV$ we are 
interested here, the neutron mean-free-path is nearly an order of 
magnitude larger than the corresponding interaction lenght for neutrons with 
the $2.73~K$ BBR in the photoproduction of pions. We just remind these 
$\bar{\nu}_e$s from neutron decay because, even if quite rare, they require a 
primordial proton energy lower than the proposed channels. Indeed, the main 
artery for the UHE neutrino production is the photopion multiproduction by 
protons: $p\gamma_{BBR}\rightarrow p+N\pi$, $p\gamma_{BBR}\rightarrow n+N\pi$ 
with $N\geq 2$. The neutron, being itself a proton 
secondary, will begin its chain with a degraded energy and, then, with a less 
favourable rate for the $\pi^\pm$ production.}. Hence, one 
could consider the W exchange interactions $\nu_e\bar{\nu}_r$ (or 
$\bar{\nu}_e\nu_{r}$), and in particular $\nu_e\bar{\nu}_{\tau_r}$, 
$\bar{\nu}_e\nu_{\tau_r}$, as a source of UHE electron secondaries. In 
principle, such UHE electrons may convert half of their energy into photons 
by Inverse Compton Scattering (ICS) on BBR. This reaction, 
being so short-cut, is very efficient in UHE photon production. The ICS 
in galactic halo, while being dominant up to $E_e\sim m_e^2/(h\nu_{opt})\sim 
2\cdot 10^{11}~eV$ over the competitive synchrotron radiation, is no longer 
ruling at very high energies. Indeed, since for $E_e > m_e^2/(h\nu_{opt})$ the 
Klein-Nishina cross section decreases linearly with the electron energy, the 
synchrotron radiation losses (and the corresponding interaction lenghts) are 
larger (smaller) by 6 orders of magnitude than the ICS ones (since synchrotron 
interaction works at $\it{Thomson}$ constant regime), for UHE electrons whose 
energy is $E_e\gg m_e^2/(h\nu_{BBR})\sim 4\cdot 10^{14}~eV$, and in 
particular for $E_e\sim 10^{21}~eV$. Therefore, we must expect only an 
associated parassite 
electromagnetic shower at the characteristic synchrotron radiation energy 
$E_\gamma\simeq 1.6\cdot 10^{16}~eV~(E_e/10^{21}~eV)^2$ at lower but 
significant flux of energy with respect to the UHECR, even in the same 
direction, but at delayed time. Moreover, because of these synchrotron 
radiation losses, the interaction lenght for synchrotron radiation, for the 
electrons inside our galaxy ($B_G\sim 3\cdot 10^{-6}~G$), is reduced to 
$\lambda_e\simeq 120~pc$. Therefore, the probability that such an electron is 
the progenitor of the observed signal at the Fly's Eye detector is negligible. 
In particular, this probability is $P_e\simeq 
\sigma_{\nu\nu_r} n_{\nu_r}\lambda_e \eta_n\sim 3\cdot 10^{-8}$ where 
$\eta_n\sim (10^{20}~eV/E_n)$ is the efficiency for beta-decay. A more 
interesting probability is obtained if we consider the electrons coming via 
the $t$-channel from the tau decay. In this case we get 
$P_\tau\sim 10^{-5}$, where now $\lambda_e\simeq l_g$, and 
$E_p\sim 2.2\cdot 10^{22}~eV$ for the initial proton energy.  

\section{The chain reactions leading to the final protons}

We now consider the reactions giving rise to final protons. This 
analysis has a particular importance because local effects could 
disfavour a shower initiated by a 320 EeV photon. In fact, the interaction 
between the UHE photon and the virtual photons of the stationary geomagnetic 
field leads to an electromagnetic cascade whose maximum is not in agreement 
with the observed data of the Fly's Eye event \markcite{bhg} (Burdman, Halzen, 
Gandhi, 1997).\newline
So, let us examine the steps leading to the final protons and summarized in 
tables 4,5. The first three processes in the two chains are the same as for 
the final photon 
production. The photoproduction of pions creates charged pions whose decay 
generates UHE neutrinos able to reach the galactic halo filled up with relic 
$\nu$s. Now, the $\nu\nu$ interactions either occur via a $Z^*$ exchange 
(table 4) or can create a $W^+W^-$ pair (table 5). Both the $Z^*$ and the 
$W$ can then undergo a subsequent hadronic decay (table 5). 
In the first case, at the relevant $Z^*$ center of mass energy, one gets an 
average of 2 protons in each hadronic final state, while from the on-shell $W$ 
hadronic decay one expects on average just one proton. As in the previous 
section, we derive the corresponding probability 
and initial proton energy. The main difference with respect to the similar 
photon channels (table 1,2,3) is the lack of a $t$-channel because of the 
absence of $\tau$ decays into protons. We also note the more 
promising role of the $W^+W^-$ channel over the $Z^o$ one.\newline 
Let us now compare the results for the photon and proton chains. 
The Z channel for photons has a better probability but requires an higher 
initial energy than the analogous one for protons. These differences are 
related to the fact that the protons are less abundant than the pions, that, 
moreover, 
must still decay into photons. So, the pions need an higher center of mass 
energy, but the consequent depletion of the cross section is balanced by the 
larger 
multiplicity. A similar behaviour can be found in the $W^+W^-$ channel for 
photons and protons. The difference in the initial energy is just the factor 
two that stems 
from the pion decay into photons, while the difference in the probability is 
related again to the different multiplicity.\newline
As we further discuss in the conclusions, the $W^+W^-$ channel into protons 
could indeed give a solution to the 320 EeV event. The necessary initial proton 
energy ($\sim 7.3\cdot10^{23}$) is quite high, but the probability 
($\geq 1.2\cdot 
10^{-3}$) is at least 120 times more favourable than the one required for the 
direct travel of a proton from the source to the Earth. Moreover, and most 
important, such a complicated sequence of reactions explains the embarrassing 
absence of the expected hundreds of cosmic ray signals at EeV energies that 
must be present for a direct proton propagation.

\section{Conclusions}

As summarized in tables 1,2,3 for the final photons, and in tables 4,5 for the 
final protons, the probabilities and primordial 
energies for proton and neutron chains for the $W$, $Z$ and 
$WW$ channels, are able to give an extragalactic solutions to the UHE 
puzzle. The neutrinos can be the $\it{ambassadors}$ of cosmic 
energetic sources whose energies are finally converted by the relic, massive 
neutrino halo $\it{calorimeter}$ into UHE photons or protons. The 
approximate lower bound on the total probability and the needed 
initial proton energy, for the different channels, and for a relic neutrino 
mass $m_\nu \sim 10~eV$, are

$$
\begin{array}{llcl}
table~1~(\gamma): & P_{tot}^W\geq 10^{-3} & & E_p^W\simeq 4.4\cdot 
10^{22}~eV\nonumber\\
table~2~(\gamma): & P_{tot}^Z\geq 6.2\cdot 10^{-3} \left( \frac{m_\nu}
{10 eV} \right)^{-1} & & E_p^Z\simeq 3\cdot 10^{24}~eV\nonumber\\
table~3~(\gamma): & P_{tot}^{WW}\geq 2\cdot 10^{-2}\left( \frac{m_\nu}
{10 eV} \right)^{-1} & & E_p^{WW}\simeq 1.8\cdot 10^{24}~eV\nonumber\\
table~4~(p): & P_{tot}^Z\geq 5.1\cdot 10^{-4} \left( \frac{m_\nu}{10 eV}
\right)^{-1} & & E_p^Z\simeq 10^{24}~eV\nonumber\\
table~5~(p): & P_{tot}^{WW}\geq 1.2\cdot 10^{-3} \left(\frac{m_\nu}
{10 eV} \right)^{-1} & & E_p^{WW}\simeq 7.3\cdot 10^{23}~eV\nonumber
\end{array}
$$
We indicated just the main neutrino mass dependence in the four last 
probabilities, while the exact one is a more complicated function. 
This analysis shows that the $W^\pm$ channel for photons and the $W^+W^-$ 
channel for protons give the most reasonable combination 
of the total probability and initial proton energy. These probability values 
are at least three order of magnitude above the corresponding ones for a 
$\it{direct}$ neutrino interaction on the 
terrestrial atmosphere. The required primordial sources may be safely located 
at any cosmic distance escaping the GZK cutoff. The Seyfert galaxy 
MCG 8-11-11, which is very close to the arrival direction of the 320 EeV 
shower and located at a redshift z=0.0205 ($D\simeq 70~Mpc~H_{100}^{-1}$), 
could be a very natural candidate. 
Its large observed luminosity in low-energy gamma ($L_\gamma\sim 7\cdot 
10^{46}~erg~s^{-1}$) is of the order of magnitude needed to explain the UHE 
energetics within our present scheme. 
Indeed, the total energy needed for 
any (spherical) source at a distance D to give rise to the 320 EeV event in 
our approach is: $E_s=(E_p\cdot 4\pi D^2)/(A\cdot P)$, where A is the Fly's Eye 
detector area [$\sim (30~Km)^2$] and P is the probability of any given 
channel. The corresponding power a source needs to get a rate of just one 
event a year, for the W, Z and WW channels, respectively, and for the most 
conservative value of the $\nu\nu$ interaction probability, is 
$$
\begin{array}{lcl}
table~1~(\gamma): & & \dot{E}_s^W\sim 2.2\cdot 10^{47} (D/100~Mpc)^2~erg~
s^{-1}\nonumber \\
table~2~(\gamma): & & \dot{E}_s^Z\sim 2\cdot 10^{48} (D/100~Mpc)^2~erg~
s^{-1}\nonumber \\
table~3~(\gamma): & & \dot{E}_s^{WW}\sim 3.7\cdot 10^{47} (D/100~Mpc)^2~erg~
s^{-1}\nonumber\\
table~4~(p): & & \dot{E}_s^Z\sim 8.1\cdot 10^{48} (D/100~Mpc)^2~erg~
s^{-1}\nonumber\\
table~5~(p): & & \dot{E}_s^{WW}\sim 2.5\cdot 10^{48} (D/100~Mpc)^2~erg~
s^{-1}\nonumber
\end{array}
$$
These values may be one or two order of magnitude overestimated, if the 
relic neutrino clustering is more efficient. Anyway, they are 
already comparable to the MeV observed power from MCG 8-11-11. 
The energetic and directionality $\it{resonance}$ toward this 
source and the quite natural hypothesis that at least a (tau) neutrino mass 
falls in the range of a few tens eV, as expected in HCDM standard 
cosmological model, seem to favour our solution of the UHECR puzzle. 
Nevertheless, we believe 
that more theoretical and experimental investigations are needed, that may 
lead to more convincing evidences of an extended dark neutrino halo, and 
possibly even to an indirect estimate of the neutrino mass.

\begin{center}
{\bf Acknowledgements}
\end{center}

\noindent
We thank Dr. Dario Grasso for bringing (Roulet 1993) to our attention.\newline 
We also thank Dr. A.Aiello, Dr. R.Conversano and M.Grossi for their careful 
reading of the manuscript.

\clearpage

\figcaption{The total cross sections for the indicated 
processes as function of the center of mass energy (for a relic neutrino mass 
$m_\nu = 10 eV$)\label{fig1}}

\clearpage

Tables caption:--- The five tables summarize on each column respectively: the 
kind of reaction, the corresponding probability, the consequent averaged
multiplicity of the final products, the secondary averaged energy. To give an
example we refer to table 1 but the same explanation is valid for the other
tables also. Reaction 1a) shows the photopion production of primary protons
of energy $E_p$ onto BBR photons either for a final $p$ or $n$ creation plus
a couple of $\pi$. The corresponding probability of pion production is
pratically the unity for the cosmic distances we assumed here. The charged
pions are 2/3 of the total number so we considered a conservative value of 1.
Finally, the secondary pion escaping from this reaction has an average energy
of 1/3 of the primordial proton energy.\newline
In the reactions 2a), 2'a) we showed the splitting of the chain into two 
branches due to the generation of UHE neutrinos from the 
charged pions decay and from the secondary muons decay.\newline
In the reactions 3a), 3a') we considered the ultrahigh neutrino interaction 
onto the relic cosmic neutrino via the cross section of eq.1 and with the relic
neutrino number density discussed in the text. The probability takes into
account the value of the cross section at the center of mass energy here
involved for the two neutrinos.\newline
In the reaction 4a), 4a') the probability shown refers to the hadronic decay 
of the tau and as a consequence the final pion multiplicity is just 1.\newline
At the end of the whole chain we calculated the global probability required
for the process, for both branches, as a product of the multiplicity and the 
probability that we derived at each step. The initial needed proton energy 
$E_p$ is then derived
from the chain requiring that at least one of the two branches could give a 
photon with the known energy of 320 EeV for the final particle.\newline
The further tables 2,3,4,5 must be read in the same way.










\end{document}